\documentclass[a4paper]{article}
\usepackage{listings}
\usepackage{xcolor}
\usepackage{multirow}
\usepackage{jheppub}  
\usepackage{dcolumn}
\usepackage[english]{babel}
\usepackage[utf8x]{inputenc}
\usepackage[T1]{fontenc}
\usepackage{palatino}
\usepackage{amsmath}
\usepackage{afterpage}
\usepackage{graphicx, subcaption}
\usepackage[colorinlistoftodos]{todonotes}
\usepackage[colorlinks=true]{hyperref}
\usepackage{makeidx}
\newcommand{\be}{\begin{equation}}  
\newcommand{\ee}{\end{equation}}  
\newcommand{\bea}{\begin{eqnarray}}  
\newcommand{\eea}{\end{eqnarray}}  
\makeindex
\begin{document}

\vspace*{1.2cm}

\thispagestyle{empty}
\begin{center}
{\LARGE \bf Recent results from the TOTEM collaboration at the LHC}

\par\vspace*{7mm}\par

{

\bigskip

\large \bf Christophe Royon}

\bigskip

{\large \bf  E-Mail: christophe.royon@ku.edu}

\bigskip

{Department of Physics and Astronomy, The University of Kansas, Lawrence KS 66047, USA}

\bigskip

{\it Presented at the Workshop of QCD and Forward Physics at the EIC, the LHC, and Cosmic Ray Physics in Guanajuato, Mexico, November 18-21 2019}


\vspace*{15mm}

\end{center}
\vspace*{1mm}

\begin{abstract}
WWe describe the most recent results from the TOTEM collaboration at the LHC, namely the elastic cross section measurements at a center-of-mass on 2.76, 7, 8 and 13 TeV. No structure or resonance is
observed at high $t$ at high center-of-mass energies. A pure exponential form of 
$d \sigma/dt$ is excluded both at 8 and 13 TeV. Accessing the very low $t$ region allows measuring the $\rho$ parameter at 13 TeV. 
\end{abstract}

 
 \section{Method to measure elastic events at the LHC}

The TOTEM collaboration measured the elastic $pp \rightarrow pp$ cross sections by detecting both intact
protons in the final state and vetoing on activities in the main CMS detector. It installed sets of vertical roman pot detectors at about 147 and 220-240 meters from the interaction
point that have good acceptance in detecting intact protons in elastic interactions. The elastic event trigger requires the presence of one intact proton on each side of the interaction point on
UP-DOWN or DOWN-UP configurations. In addition to the roman pots detectors, the TOTEM collaboration installed
two inelastic telescopes called $T_1$ and $T_2$ covering respectively the region $3.1 < |\eta| < 4.7$ and $5.3 < |\eta|
<6.5$ for charged particle $p_T$ above 100 and 40 MeV, respectively. Requesting no activity in $T1$ and $T2$ allows
suppressing 92\% of inelastic background. 

Before the TOTEM measurements, it is worth noticing that the predictions of elastic cross sections were showing large differences
by orders of magnitude especially at large $t$. Some models even predicted the existence of diffractive structures at 
medium $t$ at LHC energies. Elastic events allow probing soft diffraction and Pomeron exchange at low $t$, diffractive
structures at medium $t$ and parton scattering and perturbative QCD effects at larger $t$.
 
\begin{figure}
\begin{center}
\epsfig{figure=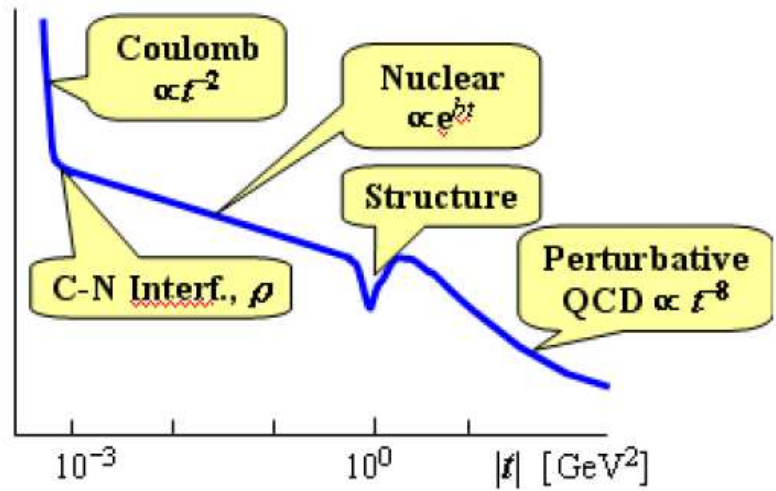,height=0.45\textwidth}
\caption{Schematic of elastic cross section measurement.}
\label{coulomb}
\end{center}
\end{figure}

\begin{figure}
\begin{center}
\epsfig{figure=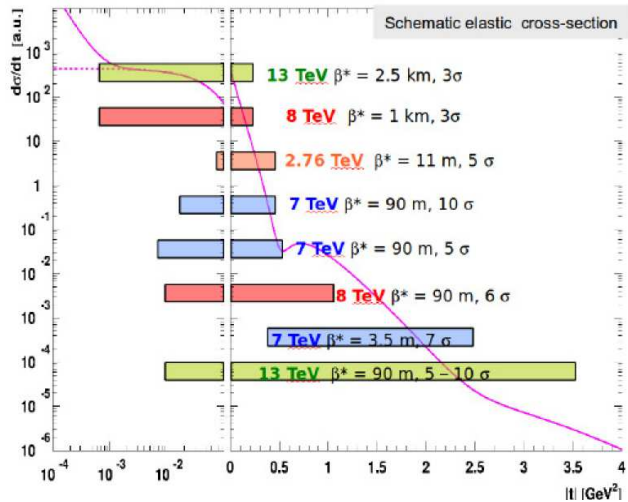,height=0.45\textwidth}
\caption{Different kinematical domains and $\beta^*$ values where the elastic cross section $d \sigma/dt$ was measured by the TOTEM experiment.}
\label{elcross}
\end{center}
\end{figure}

In Fig.~\ref{coulomb}, we display the different domains of measurement for the $d \sigma/dt$ elastic cross section.  At very low $t$, the cross section is proportional to $1/t^2$ (Coulomb region) and at higher $t$, we reach the nuclear region where the
cross section is proportional to $e^{bt}$. Between these two regions, we need to take into account the interferences between
the nuclear and Coulomb cross sections (called C-N interference in the figure or CNI) where we can measure the $\rho$ parameter,
the ratio of the imaginary part to the real part of the elastic cross section. At higher $t$, we reach the resonance or ``structure" region and at even higher $t$, the perturbative QCD region where the cross section varies as $1/t^8$. It means that one needs to access a very wide range in $t$ if one wants to measure precisely the elastic cross section.

For this sake, many different beam lattices were used at the LHC as shown in Fig.~\ref{elcross}. Accessing very low $t$ values means going to very low angles of scattered protons.  High values of $\beta^*$ such as 2.5 km at 13 TeV or 1 km at 8 TeV allow reaching low values of $t$, and the CNI region.  The accumulated luminosity is small at high $\beta^*$ and higher values of $\beta^*$ are used to cover the higher $t$ region
in order to collect more luminosity since the cross falls exponentially as a function of $t$.

Technically, in oder to measure the elastic cross section as a function of $t$, one needs to count the number of elastic protons as a function
of $t$ in the TOTEM roman pot detectors with a precision better than 2 or 3\%. Measuring the cross section at very low $t$ in the CNI region
requires going down to proton scattering angles of 3.5$\mu$rad, or $t\sim$6.5 10$^{-4}$ GeV$^2$. This requires the special high $\beta^*$ beam optics as we mentioned already, detectors at $\sim$ 1.5 mm from LHC beam axis,  spatial resolution well below 100 $\mu$m and
no significant inactive edge ($< 100 \mu$m).

\section{TOTEM measurements}

\begin{figure}
\begin{center}
\epsfig{figure=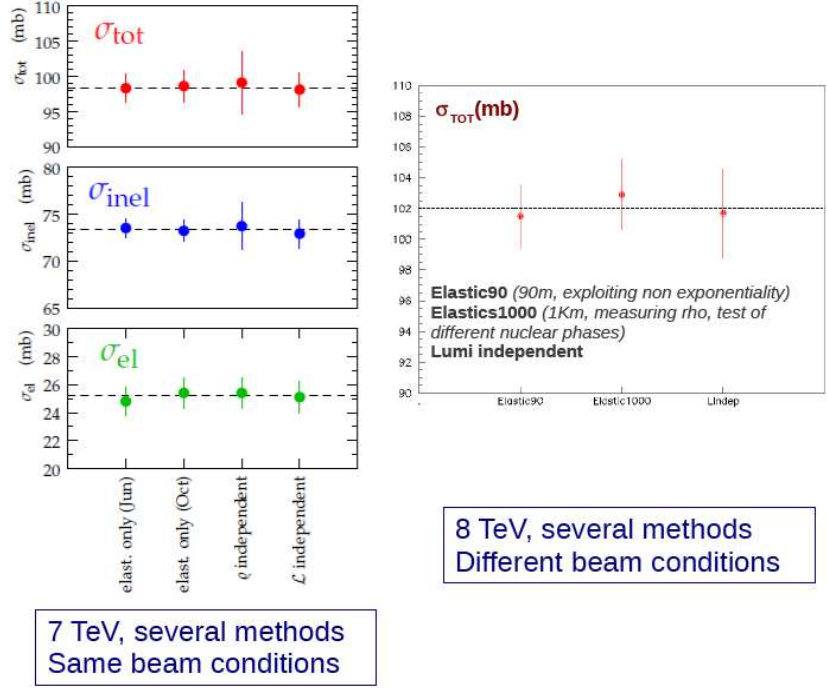,height=0.65\textwidth}
\caption{Measurements of the elastic, inelastic and total cross sections at a center-of-mass energy of 7 and 8 TeV using the different methods described in the text.}
\label{crossb}
\end{center}
\end{figure}

\begin{figure}
\begin{center}
\epsfig{figure=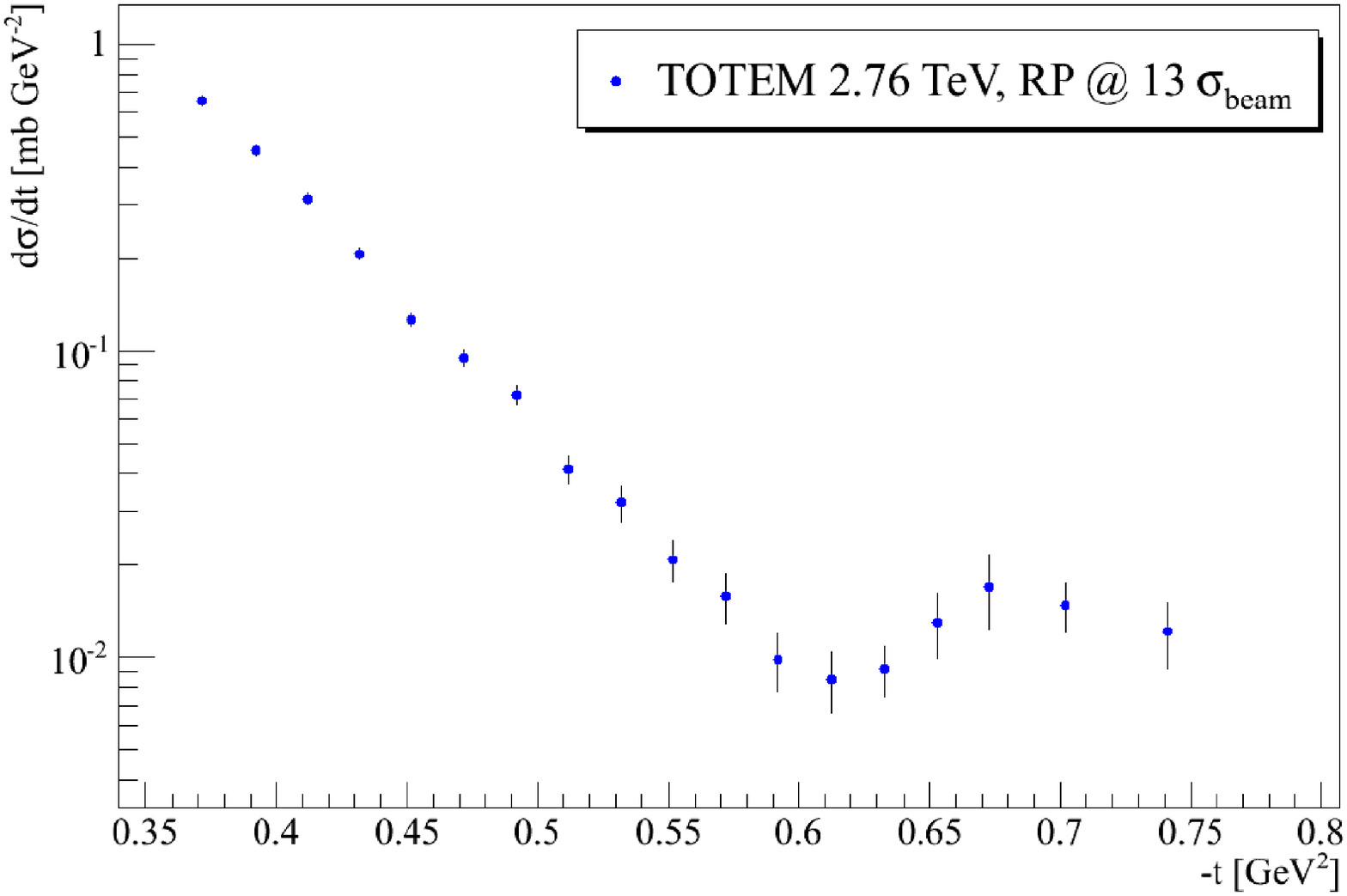,height=0.4\textwidth}
\epsfig{figure=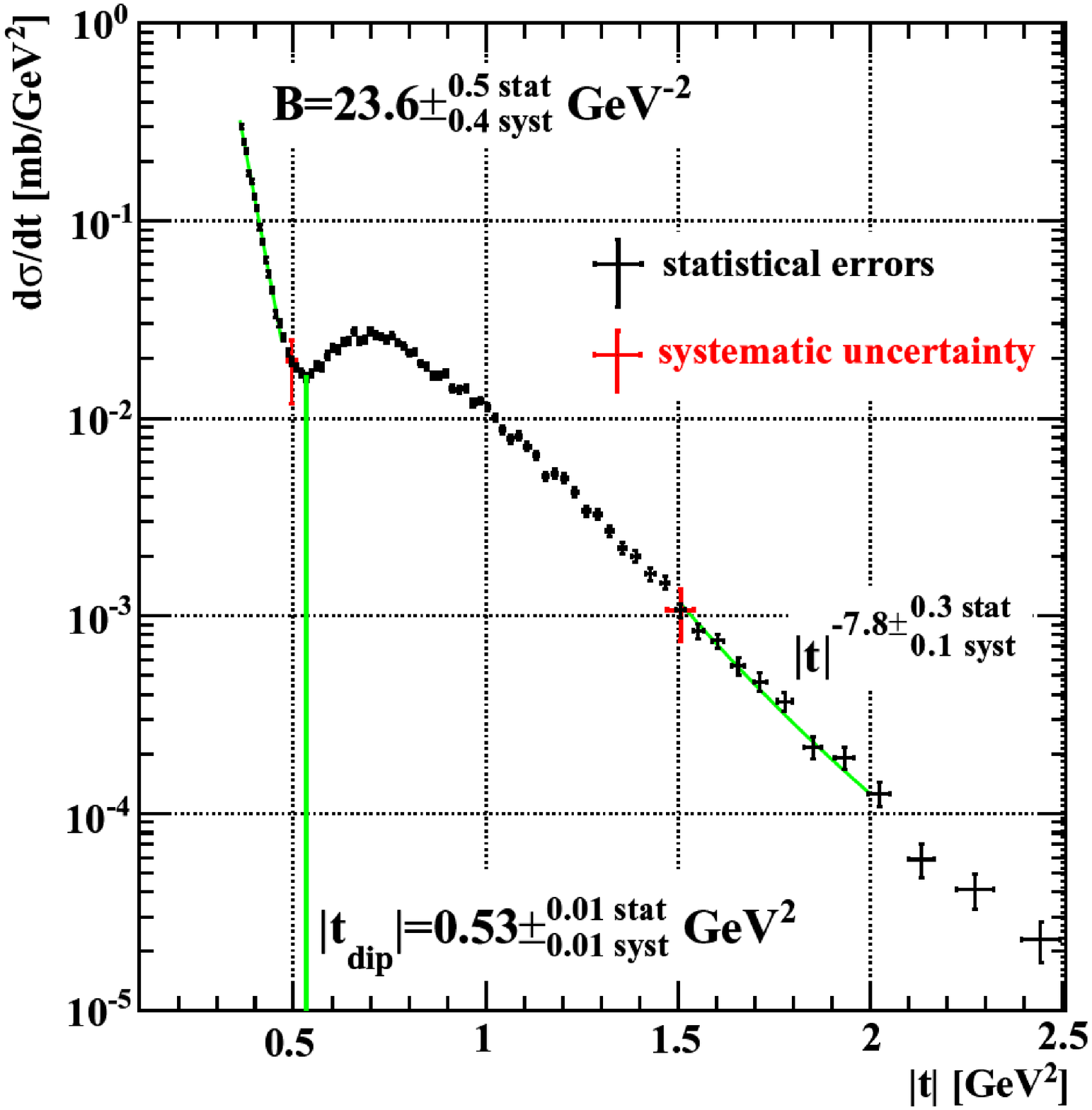,width=0.4\textwidth} 
\epsfig{figure=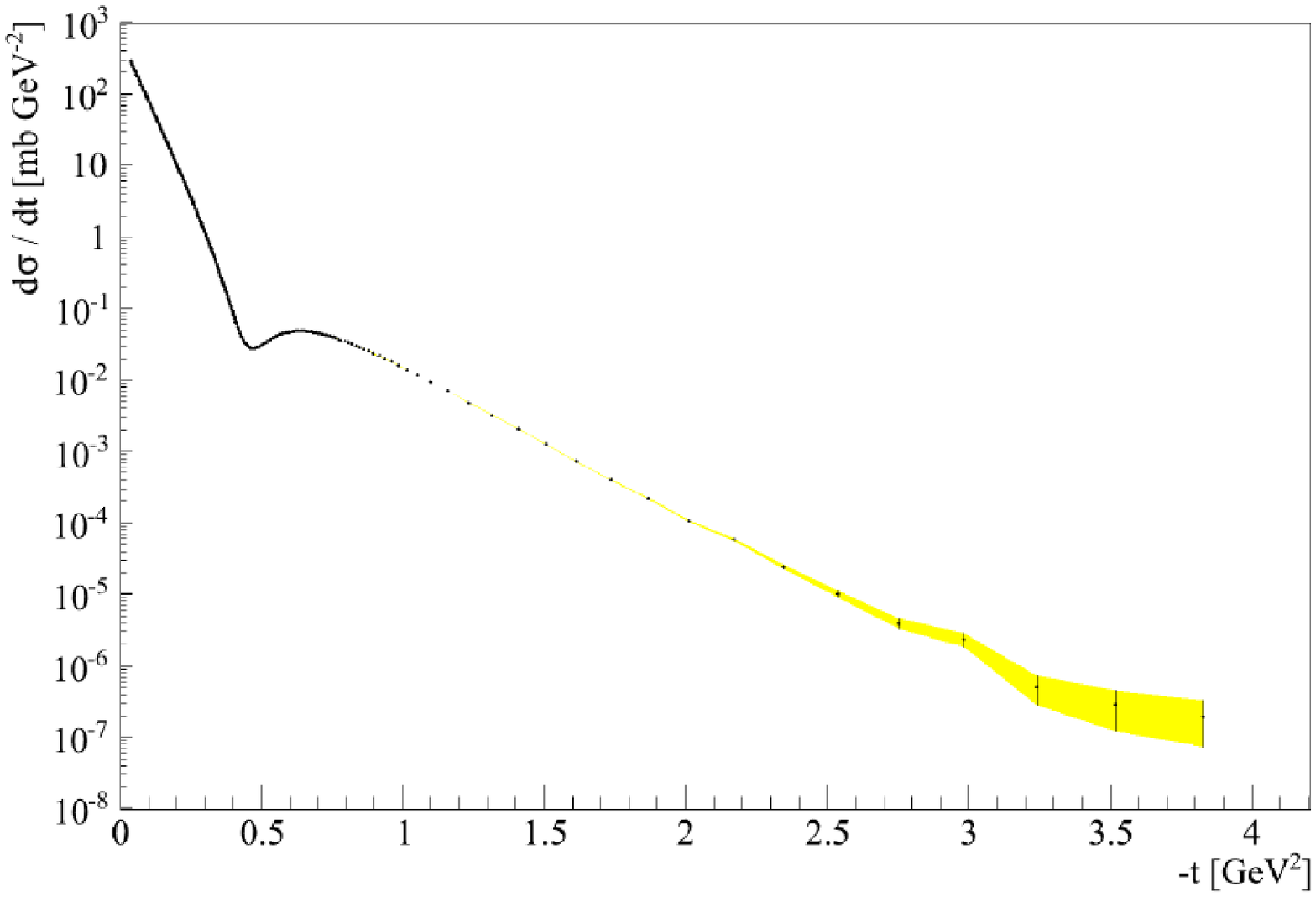,height=0.4\textwidth}
\caption{Elastic $d \sigma/dt$ cross sections measured by the TOTEM collaboration at center-of-mass energies of 2.76, 7 and 13 TeV.}
\label{elastic}
\end{center}
\end{figure}

\begin{figure}
\begin{center}
\epsfig{figure=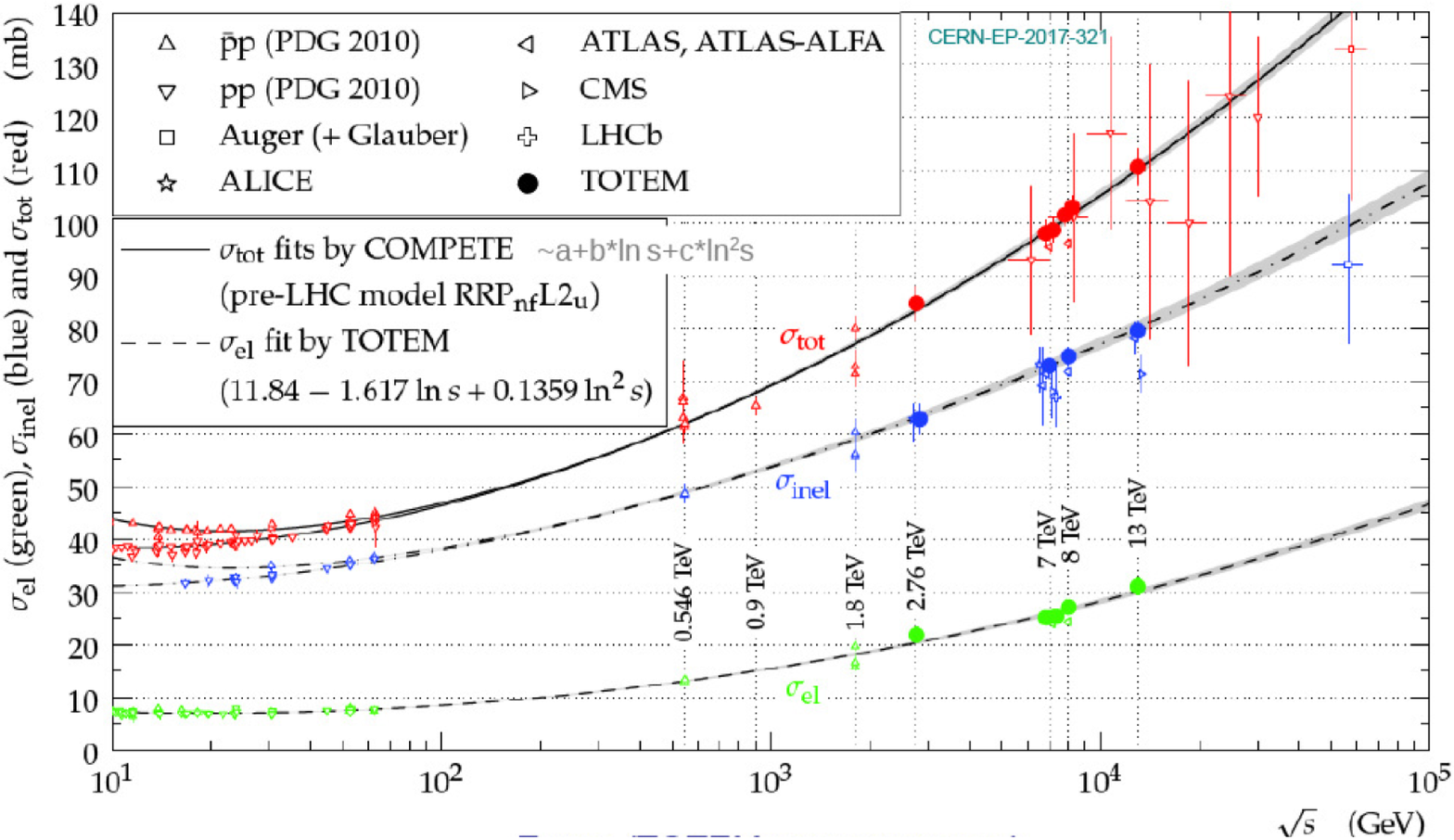,height=0.5\textwidth}
\caption{Elastic, inelastic and total cross sections measured at the LHC by the ALICE, ATLAS, CMS, LHCb and TOTEM
collaborations compared to previous measurements and to fits performed by the TOTEM and COMPETE collaborations}
\label{totalcross}
\end{center}
\end{figure}

\subsection{Total cross section and $\rho$ measurements}

The  number of elastic events $N_{el}$ is measured by tagging the protons in the roman pot detectors while vetoing on the $T_1$ and $T_2$ telescopes, while the number of inelastic events $N_{intel}$ is counted using the activity in $T_1$ and $T_2$. The method to extract the total cross sections uses the optical theorem
\begin{eqnarray}
L \sigma_{tot}^2 &=& \frac{16 \pi}{1 + \rho^2} (dN_{el}/dt)_{t=0} \\
L \sigma_{tot} &=& N_{el} + N_{inel}
\end{eqnarray}
where $L$ is the integrated luminosity, $\rho$ the ratio of the real to the imaginary part of the total cross section, and $\sigma_{tot}$ the total cross section. It leads to three different methods to extract $\sigma_{tot}$
\begin{itemize}
\item The luminosity independent measurement
\begin{eqnarray}
\sigma_{tot} = \frac{16 \pi}{(1 + \rho^2)}
\frac{(dN_{el}/dt)_{t=0}}{(N_{el}+N_{inel})} 
\end{eqnarray}
\item The luminosity dependent measurement (elastic only)
\begin{eqnarray}
\sigma_{tot}^2 = \frac{16 \pi}{(1 + \rho^2)} \frac{1}{L} (dN_{el}/dt)_{t=0}
\end{eqnarray}
\item The $\rho$ independent measurement
\begin{eqnarray}
\sigma_{tot} = \sigma_{el} + \sigma_{inel}
\end{eqnarray}
\end{itemize}

At very low $t$, in the CNI region, $d\sigma/dt$ can be rewritten as
\begin{eqnarray}
\frac{d \sigma}{dt} \sim |A^C + A^N ( 1 -\alpha G(t))|^2 
\end{eqnarray}
where $A^C$ and $A^N$ are the Coulomb and Nuclear amplitudes, and $G(t)$ the interference term. 
The differential cross section is sensitive to the phase of the
nuclear amplitude, and in the CNI region, both the modulus and the phase of the nuclear 
amplitude can be used to determine
\begin{eqnarray}
\rho = \frac{Re(A^N(0))}{Im (A^N(0))}  
\end{eqnarray}
where the modulus is constrained by the measurement in the hadronic region and the
phase by the $t$ dependence.

The three independent methods to measure the elastic, inelastic and total cross sections are in good agreement, and as an example, the results are shown in Fig.~\ref{crossb} for a center-of-mass energy of 7 and 8 TeV. In addition, the elastic, inelastic and total cross section measurements performed by all LHC experiments are shown in Fig.~\ref{totalcross}. The total cross section at high energies is compatible within uncertainties with previous results from
cosmic ray experiments.  The increase of $\sigma_{el}/\sigma_{tot}$ with energy is confirmed
at LHC energies up to 13 TeV. A recent measurement of the total cross section from LHCb at 13 TeV is also compatible with
TOTEM results~\cite{lhcbnewcross}. It is worth noticing that there is a discrepancy of about 1.9$\sigma$ at 8 TeV between ATLAS and TOTEM results, and new measurements by ATLAS at 13 TeV will be of great interest in order to check if this discrepancy persits at higher center-of-mass energies. As we will see in the following, $\rho$ was also measured using the $\beta^*=$2500 m data.

\subsection{Measurements of $d\sigma/dt$ and implications}
The $t$-dependent measurements of the elastic cross section
for center-of-mass energies of 2.76, 7 and 13 TeV are shown in Fig.~\ref{elastic}. We note the presence of a dip and a maximum 
towards $|t| \sim 0.5-0.6$ GeV$^2$ at all center-of-mass energies~\cite{total}. The dip position in $t$ also decreases with increasing $\sqrt{S}$. The other noticeable result is that there
is no structure at high $t$ at high center-of-mass energies as shown in Fig~\ref{elastic} contrary to what some parameterizations assumed before the LHC era. 

\begin{figure}
\begin{center}
\epsfig{figure=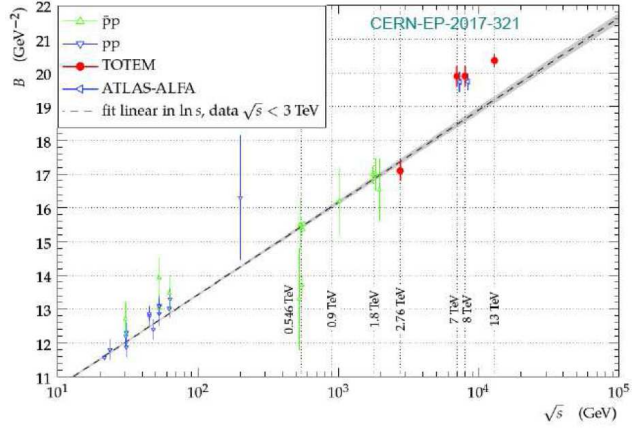,height=0.40\textwidth}
\caption{$B$-slope of the elastic cross section as a function of center-of-mass energy.}
\label{bslope}
\end{center}
\end{figure}

\begin{figure}
\begin{center}
\epsfig{figure=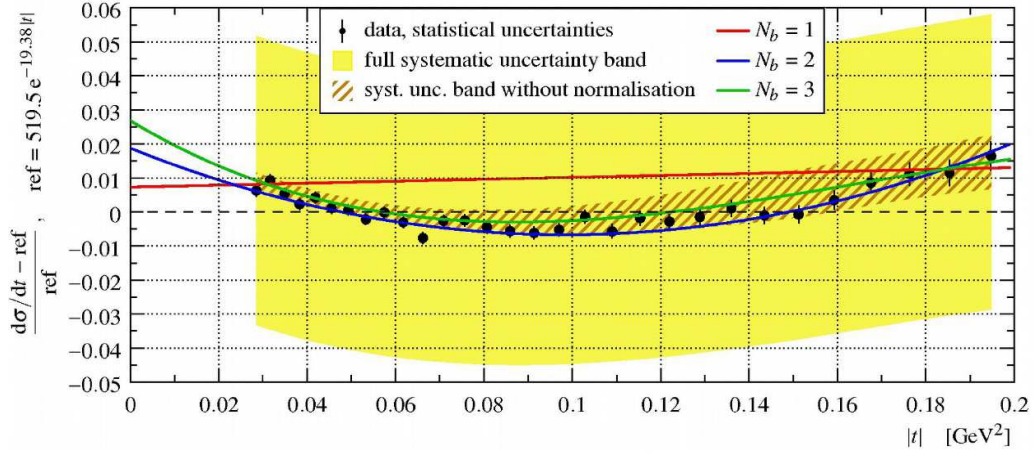,height=0.40\textwidth}
\caption{Non-exponential behavior of TOTEM elastic data at a center-of-mass energy of 8 TeV
}
\label{nonexp}
\end{center}
\end{figure}

The $B$-slope of the elastic $d\sigma/dt$ cross section is shown in Fig.~\ref{bslope}. It is found to be larger  at 13 TeV than at lower center-of-mass energies and a 
linear behavior ($ln s$) is compatible up to $\sqrt{s}<3$ TeV, but it becomes non-linear at higher energy.
An attempt to fit the elastic cross section $d \sigma/dt$ using a usual simple exponential fit at low $t$ was performed
\begin{eqnarray}
d \sigma/dt = A \exp (-B(t) |t|)
\end{eqnarray}
and different polynomial fits of $B(t)$ were used
\begin{itemize}
\item 1st order polynomial fit ($N_b=1$):  $B=b_1$, seen as a reference
\item 2nd order polynomial fit ($N_b=2$):  $B=b_1+b_2t$
\item 3rd order polynomial fit ($N_b=3$): $B=b_1+b_2t+b_3 t^2$ 
\end{itemize}
A pure simple exponential form ($B=b_1$) is excluded at 7.2$\sigma$ with 8 TeV data as shown in Fig.~\ref{nonexp}, where we
display the relative difference with the ($N_b=1$) fit. The uncorrelated uncertainties are in red and the full systematic in yellow. Similar results were found using 13 TeV data.

\begin{figure}
\begin{center}
\epsfig{figure=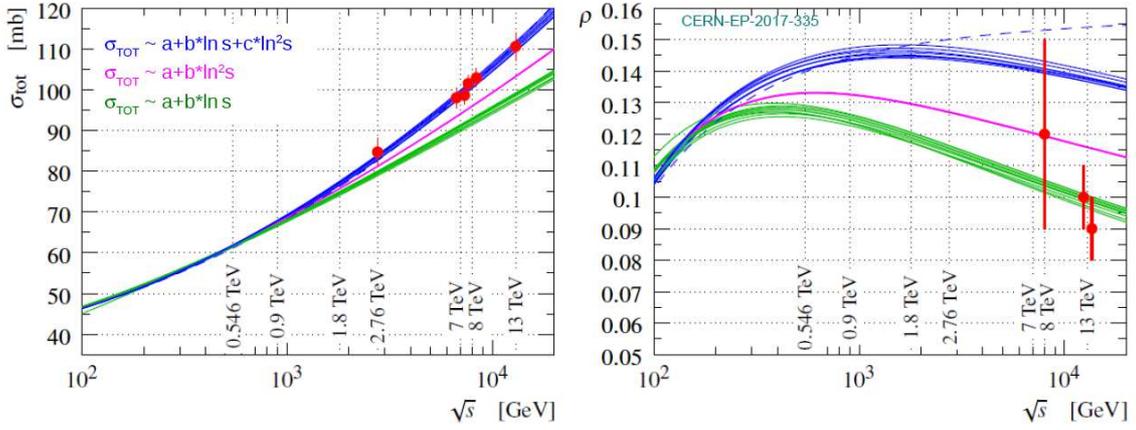,height=0.40\textwidth}
\caption{Measurements of the total cross sections and the $\rho$ parameter as a function of $\sqrt{s}$ compared to 
fits using linear, quartic and a combination of linear and quartic terms.}
\label{rho}
\end{center}
\end{figure}

\begin{figure}
\begin{center}
\epsfig{figure=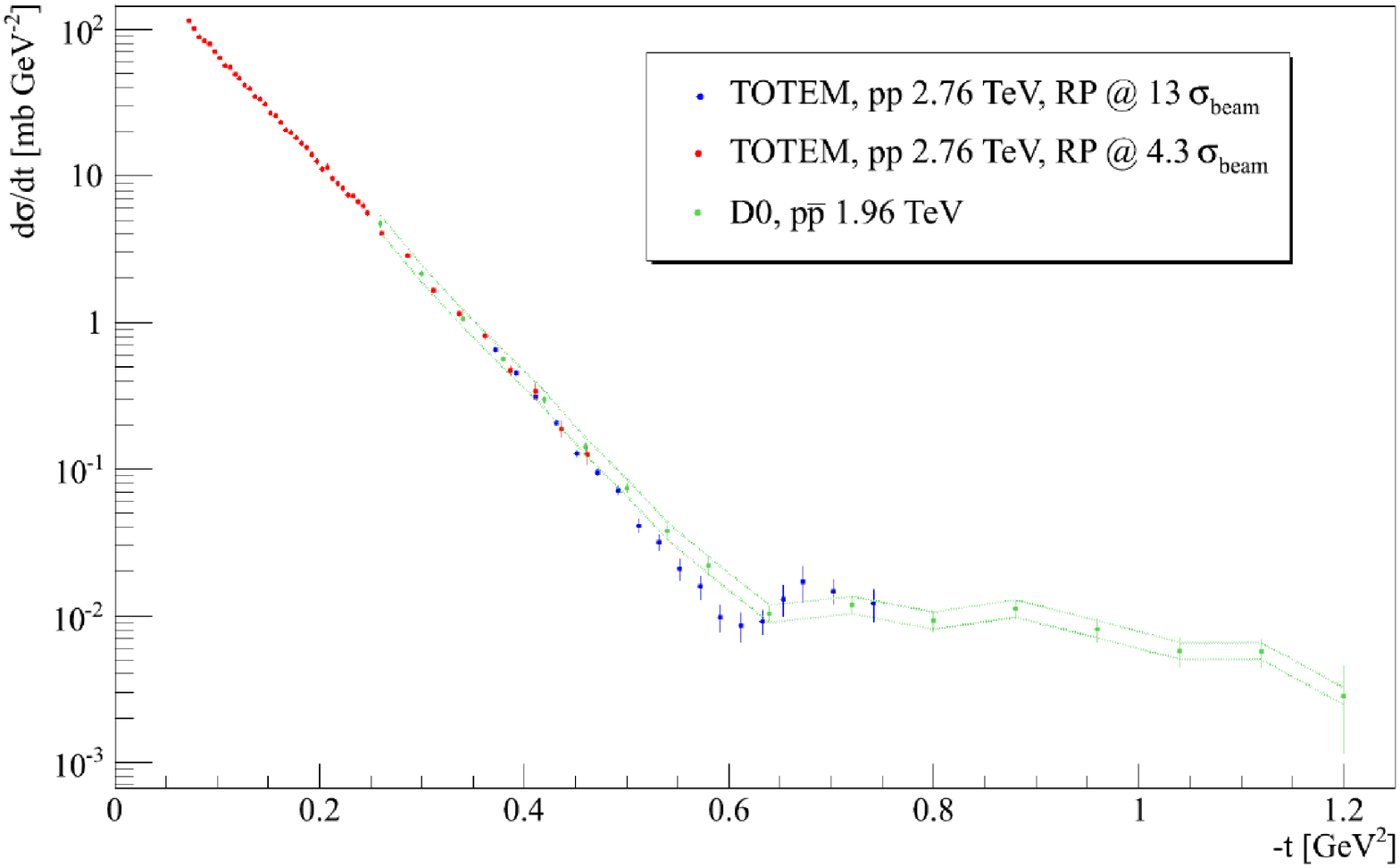,height=0.45\textwidth}
\caption{Comparison between the $pp$ TOTEM cross section measurement at 2.76 Tev and the $p \bar{p}$ measurement from D0 at 1.96 TeV.}
\label{d0}
\end{center}
\end{figure}

The very precise measurement of the total cross section at 13 TeV down to very 
low $|t|$ allows to measure with high accuracy the $\rho$ parameter, $\rho=0.09\pm 0.01$~\cite{totemrho}. The values of the total cross section $\sigma_{tot}$ and the
$\rho$ parameter as a function of $\sqrt{s}$ are shown in Fig.~\ref{rho}. They are compared to a linear, or a quartic fit
to $\sigma_{tot}$ as a function of $\sqrt{s}$ as well as to a combined linear and quartic fit. $\sigma_{tot}$ data 
clearly favor the combined fit whereas the $\rho$ measurement at 13 TeV favors the linear fit. The difference between these
two observables can be interpreted as an additional colorless exchange not introduced in these simple fits, namely
the Odderon or 3-gluon exchanges. In order to have better evidence for the existence of the Odderon, it is useful
to compare directly $pp$ and $p \bar{p}$ cross sections. $pp$ and $p \bar{p}$ data from the TOTEM and D0~\cite{d0cross}
are shown in Fig.~\ref{d0}. Even if data were not taken at the same $\sqrt{s}$, it is worth noticing that the $pp$ data
at 2.76 TeV show a dip and maximum  whereas $p \bar{p}$ data do not show such a structure. The idea is to identify characteristic features of
$pp$ data and compare them with $p \bar{p}$ results, for instance the dip and bump that are not visible in $p \bar{p}$ interactions. While some
quantitative studies are still being performed by the D0 and TOTEM collaborations, it is clear that a natural
explanation for the difference between both colliders is due to Odderon exchanges.

\section{Conclusion}
In this article, we reviewed the most recent results from the TOTEM collaboration. Accessing low values of $t$ allowed the TOTEM collaboration to measure the elastic $d\sigma/dt$ cross section at 2.76, 7, 8, and 13 TeV with unprecedented precision. The $B$-slope of the elastic cross section is found to be larger at 13 TeV. The dip position in $d\sigma/dt$ decreases with $\sqrt{S}$ and no structure or resonance is
observed at high $t$ at high center-of-mass energies. The high precision on data allowed to show that a pure exponential form of 
$d \sigma/dt$ is excluded both at 8 and 13 TeV. Going to very low $t$ allowed to measure $\rho$ at 13 TeV and $\rho$ and $d\sigma/dt$ cannot be easily described within the same model. This can be interpreted as the possible existence of the Odderon. The observation of a potential difference between $pp$ and $p \bar{p}$ elastic cross sections  would lead to a clean observation of the Odderon.
 
\section*{Acknowledgements}

The author thanks the support from the Department of Energy, contract DE-SC0019389.

\end{document}